\documentstyle[prd,aps,preprint]{revtex}
        		
\newcommand\ee{\end{equation}}
\newcommand\be{\begin{equation}}
\newcommand\eea{\end{eqnarray}}
\newcommand\bea{\begin{eqnarray}}

\newcommand\lsim{\mathrel{\rlap{\lower4pt\hbox{\hskip1pt$\sim$}}
    \raise1pt\hbox{$<$}}}
\newcommand\gsim{\mathrel{\rlap{\lower4pt\hbox{\hskip1pt$\sim$}}
    \raise1pt\hbox{$>$}}}

\def\dslash{\not{\hbox{\kern-2pt $\partial$}}}
\def\Dslash{\not{\hbox{\kern-4pt $D$}}}
\def\Oslash{\not{\hbox{\kern-4pt $O$}}}
\def\Qslash{\not{\hbox{\kern-4pt $Q$}}}
\def\pslash{\not{\hbox{\kern-2.3pt $p$}}}
\def\kslash{\not{\hbox{\kern-2.3pt $k$}}}
\def\qslash{\not{\hbox{\kern-2.3pt $q$}}}
 \newtoks\slashfraction
 \slashfraction={.13}
 \def\slash#1{\setbox0\hbox{$ #1 $}
 \setbox0\hbox to \the\slashfraction\wd0{\hss \box0}/\box0 }

\def\eeq{\end{equation}}
\def\beq{\begin{equation}}

\begin{document}

\draft
\tighten
\preprint{
OUTP-97-43-P}

\title{More about Electroweak Baryogenesis\\ in the Minimal Supersymmetric Standard Model}
\author{Antonio Riotto$^{\P}$}

\address{Department of Physics, Theoretical Physics, University of Oxford\\
1 Keble Road, Oxford OX1 3NP, United Kingdom}
\date{September 1997}
\maketitle

\begin{abstract}
We compute the baryon asymmetry generated at the electroweak phase transition by the Higgs scalar sector of the minimal supersymmetric standard model. Because of large enhancement effects from low momentum modes, Higgs particles may be responsible for the observed baryon asymmetry even though CP-violation in the Higgs sector only appears at the one-loop level. We also discuss  the approximations made  in the  analysis and suggest possible improvements.

\vskip4.cm
\begin{flushleft}
$^{\P}$ Advanced PPARC Fellow. From December 1997 through November 1999 on leave of absence at the CERN Theory group as CERN Fellow.
\end{flushleft}

\end{abstract}
\newpage
\baselineskip=20pt

{\bf 1.}~~The Standard Model (SM) fulfills all the requirements for  a successful   
generation of the baryon number at the electroweak scale \cite{reviews}  due to the presence of  
baryon number violating processes which also impose severe constraints on  
models where the baryon asymmetry is created at very high energy scales  
\cite{anomaly}. Unfortunately, the electroweak phase transition is at best weakly  
first order in the SM \cite{transition}  meaning that the baryon asymmetry  
generated during the transition would be subsequently erased by unsuppressed  
sphaleron transitions in the broken phase. 
Therefore, if one is willing to pursue the idea of electroweak baryogenesis,  
some new physics at the weak scale must be  called for. The most promising and  
well-motivated candidate seems to be supersymmetry (SUSY).  Electroweak  
baryogenesis in the framework of the Minimal Supersymmetric Standard Model  
(MSSM) has  attracted much attention in the past years with 
particular emphasis on the strength of the phase transition ~\cite{early1,early2,early3} and  
the mechanism of baryon number generation \cite{nelson,noi,ck}.

Recent  analytical \cite{r1} and  lattice  
computations  \cite{r2} have  revealed   that the phase transition can be sufficiently strongly  
first order if the lightest  stop is not much heavier than the top quark, the
ratio of the vacuum expectation values of the two neutral Higgses $\tan\beta$  
is smaller than $\sim 4$ and the lightest Higgs  is  lighter than about 85 GeV. 
Moreover, the MSSM has the bonus that additional sources  
of CP-violation may be present  besides the CKM matrix phase. These new phases are essential  for the generation of the baryon number since  large  
CP-violating sources may be  locally induced by the passage of the bubble wall separating the broken from the unbroken phase.   Baryogenesis is fueled  when transport properties allow the CP-violating  
charges to efficiently diffuse in front of the advancing bubble wall where
anomalous electroweak baryon violating processes are not suppressed.
The new phases   
appear    in the soft supersymmetry breaking  
parameters associated to the stop mixing angle and to  the gaugino and  
neutralino mass matrices; large values of the stop mixing angle
are, however, strongly restricted in order to preserve a
sufficiently strong first order electroweak phase transition.
Therefore, an acceptable baryon asymmetry from the stop sector
may only be generated through a delicate balance between the values
of the different soft supersymmetry breaking parameters contributing
to the stop mixing parameter, and their associated CP-violating
phases \cite{noi}.  On the other hand, charginos and neutralinos may be responsible for the observed baryon asymmetry if   the phase of the parameter $\mu$ is    larger than about 0.1 \cite{noi,ck}. 

Since   in the  
case in which the  bubble wall is thick, the Higgs background  
carries  a very low momentum   (of order of the inverse of the bubble  
wall width $L_w$), the production of the baryon asymmetry is enhanced if the degrees of freedom in  the  
gaugino/neutralino 
sectors are nearly degenerate in mass \cite{noi}: values of the  
phase of $\mu$  lower than 0.1 are only consistent with
the observed baryon asymmetry for values of $|\mu|$ of
order of the gaugino mass parameters. This is due to a large
enhancement of the computed baryon asymmetry for these values
of the parameters.   

The fact that  CP-violating sources  are     
most  easily built up  by the transmission of low momentum particles over a distance $L_w$ \cite{nelson,noi}  is an indication that  particles with masses much smaller than the temperature $T$ may be relevant in the process of quantum interference  leading to CP-violating currents in the bubble wall.  It is therefore fair to guess that  the nearly massless   neutral and charged  modes present in the scalar Higgs sector during the electoweak phase transition may  play a significant role  in supersymmetric electroweak baryogenesis.  This is analogous to what happens in the SM where the   barrier in the effective potential separating the two minima  at finite temperature  is  originated  by infrared effects.  

CP-violating Higgs currents  are generated by the interactions with  the bubble wall only  if some CP-violation is present in the scalar potential. This does not happen at the tree-level  because the form of the scalar potential is dictated by supersymmetry and CP-violating operators are forbidden. However, once supersymmetry is broken and 
if CP-violation is present in the  stop and in the -ino sector, 
 CP-violating Higgs operators are induced at finite temperature by  one-loop diagrams involving  stops, charginos and neutralinos \cite{cpviolation}.  CP-violating scatterings of the Higgs modes with the adavancing bubble wall are therefore expected to give rise to some baryon asymmetry.  

The goal of this Letter is to compute the final baryon asymmetry  generated by the Higgs scalar  sector. Because of the large enhancement from the infrared region,   Higgs particles   may be responsible for the observed baryon asymmetry even though the CP-violation  in  the Higgs sector only appears as a one-loop effect. The details of the computation are contained in the next section, while 
section 3 is devoted to our  conclusions and comments.  

{\bf 2.}~~To compute the CP-violating Higgs current
  we  make use of the method proposed in Ref. \cite{oe} and recently adopted  by Carena  {\it et al.} \cite{noi} to compute the stop and the -ino contribution to the baryon asymmetry.  This method  is entirely
based on
a nonequilibrium quantum field theory
diagrammatic approach. It   may be
applied
for all wall shapes and sizes of the bubble wall and it naturally
incorporates the effects of the  incoherent nature of plasma physics
on
CP-violating observables. What we need to compute is  
the temporal evolution of a classical order
parameter,
namely the CP-violating Higgs  current, with definite initial
conditions.
In this respect, the ordinary equilibrium quantum field theory at
finite temperature may not be applied, since it mainly
deals with transition amplitudes in particle reactions. Therefore, 
the  closed-time path  formalism is used,
which is a powerful Green's function
formulation for describing nonequilibrium phenomena
in field theory~\cite{chou}. 

Following \cite{nelson,noi}, we are interested in the generation of some charge which is 
approximately conserved in the symmetric phase, so that it
can efficiently diffuse in front of the bubble where baryon number
violation is fast, and non-orthogonal to baryon number,
so that the generation of a non-zero baryon charge is energetically
favoured.
A charge with these characteristics
is  the Higgs charge density $\langle J^0_H\rangle =\langle J^0_{H_1}-J^0_{H_2}\rangle$, where $J_{H_i}^0(z)= i\left(H_i^\dagger\partial^0 H_i-\partial^0
H_i^\dagger H_i\right)$ with $i=1,2$ and 
$H_1$ and $H_2$ denote the two Higgs doublets. 
The CP-violating sources $\gamma_H(z)$
(per unit volume and unit time) of the  Higgs  charge density
$J_H^0$ associated to the current $J_H^\mu$ and accumulated by the moving wall at a point
$z^\mu$ of the plasma can then be constructed from
$J_H^0(z)$~\cite{nelson} using the relation $\gamma_H(z)=\partial_0 \langle J_H^0(z)\rangle$.
This is  appropriate to describe the damping effects
originated by the
plasma interactions, but
does not incorporate any relaxation time scale
arising when diffusion and particle changing interactions are
included.
However, one can
leave aside diffusion and particle changing interactions
and account for them
independently in the rate
equations. This is a good approximation (at least for small bubble wall velocity) since, for instance,  the typical diffusion time $t_D\sim D/v_w^2$, where $D$ is the typical diffusion constant, is much larger  than any  other time scales~\cite{nelson,noi}.

The CP-violating part of the MSSM Higgs scalar potential reads
\begin{equation}
V_{{\rm CP}}=\lambda_5\left(H_1 H_2\right)^2+
\lambda_6\left|H_1\right|^2H_1 H_2+\lambda_7\left|H_2\right|^2H_1 H_2
+{\rm h.c.}
\end{equation}
As mentioned in the introduction, the coefficients $\lambda_{5,6,7}$ are zero at the tree-level, but are nonvanishing at the one-loop level at finite temperature  when interactions of the Higgs fields with charginos and neutralinos\footnote{The stop contribution is suppressed when  the left-handed stop mass is much larger than the temperature. This condition is necessary for the     phase transition to be  strong enough \cite{r1}.} are   taken into account \cite{cpviolation}. Since these coefficients appear only once supersymmetry  is broken, their values are proportional to the gaugino masses $M_{1,2}$,  and to the supersymmetric parameter $\mu$. 
The typical value for   $\lambda_{6,7}$   is about   $10^{-3}$, while $\lambda_5$ is at most $10^{-4}$ \cite{cpviolation}.    What is relevant for us is that  the $\lambda$'s are complex if the parameter $\mu$ carries a phase $\phi_\mu$, $\lambda_{5}=|\lambda_{5}|{\rm e}^{i 2\phi_\mu}$, $\lambda_{6,7}=|\lambda_{6,7}|{\rm e}^{i\phi_\mu}$. It is the presence of this phase that gives origin to the baryon asymmetry. 

The interesting dynamics for baryogenesis takes place in a region
close to or inside the bubble wall and we approximate it with an infinite plane
traveling at a constant speed $v_w$ along the $z$-axis.

The computation of the CP-violating Higgs current   is done by expanding 
$\langle J_H^\mu\rangle$ around the symmetric phase defined by $\langle H_1^0\rangle\equiv v_1(T)=0$ and $\langle H_2^0\rangle\equiv v_2(T)=0$. This is equivalent to a diagrammatic approach  where  $\langle J_H^\mu\rangle$ is expressed as a  sum  of different Feynman diagrams with increasing powers of external Higgs insertions. The external Higgs background has to be identified with the bubble wall configuration and    the expansion is only justified in a region of the parameter space for which the mean free time $\tau$ is smaller than the scale of variation of the masses, {\it i.e.} the wall thickness $L_w$. 
The reader is referred to refs.  \cite{oe,noi} for more details.
Since the coefficients of the expansion are computed in the symmetric phase, we must   deal with the resummation of the propagators of the
Higgs fields in order to deal with infrared divergencies \cite{arnold}. Indeed, to account for the 
interactions with the surrounding particles of the thermal bath,
particles must be
substituted by quasiparticles and dressed propagators are to be adopted.
 Self-energy
corrections at one- or two-loops to
the propagator modify the dispersion relations and
introduce a nontrivial damping rate $\Gamma$  due to the imaginary
contributions to the self-energy. 
In what follows we shall adopt  dressed propagators to compute the thermal
averages of the  composite Higgs  operator, which allows us us to naturally and
self-consistently
include  the effects of the  incoherent nature of plasma physics.

In the unbroken phase, the Higgs spectrum contains two complex
electrically neutral fields and two charged ones. At the tree-level, the
squared masses of one of the neutral states and one of the charged ones
are negative, since the origin of the field space becomes a minimum of
the effective potential only after inclusion of the finite temperature
corrections. The resummation can be achieved by considering the
propagators for the eigenstates of the thermal mass matrix, which has
positive eigenvalues given by
\begin{equation}
m_{h,H}^2=\frac{m_1^2(T)+m_2^2(T)\mp\sqrt{\left(
m_1^2(T)-m_2^2(T)\right)^2+4\:m_3^4(T)}}{2},
\end{equation}
where the $m_i^2(T)$ are the thermal corrected mass parameters of the effective 
potential, $m_1^2(T)\simeq m_1^2+(3/8)g^2 T^2$,
$m_2^2(T)\simeq m_2^2+(1/2)h_t^2 T^2$, while $m_3^2(T)=m_{3}^2$ ($+$ logarithmic corrections in $T$). Here $m_{1,2,3}^2$ are the $T=0$ coefficients of the operator
$|H_1|^2$, $|H_2|^2$ and $(H_1H_2)$ respectively and  $g$ and $h_t$ are the $SU(2)_L$
gauge coupling and the top Yukawa coupling.
Correspondingly, the neutral complex eigenstates are given by
\begin{equation}
\left\{
\begin{array}{ccc}
h&=& c_\theta\:H_1^{0 *} + s_\theta\:H_2,\\
H&=& -s_\theta\:H_1^{0 *} + c_\theta\:H_2,
\end{array}
\right.
\end{equation}
where  $c_\theta=\cos\theta$, $s_\theta=\sin\theta$ and $\theta$ is  the critical angle identified by the flat direction  of the effective potential around $v_1=v_2=0$ at the critical temperature. 
Analogous formulae hold for the charged eigenstates.

Following refs. \cite{oe,noi},  it is not difficult to show that  $\langle J_H^\mu(z)\rangle$ gets contributions from several one-loop Feynman diagrams which are obtained   by assigning
in all the possible ways the  space-time points $z$ and $x$ (which is the point where the external Higgs configurations are  attached to)
 on the positive
or negative time branches typical of the nonequilibrium approach \cite{chou}.  
Disregarding the $T=0$ piece, the  result is (in the plasma frame)
\begin{eqnarray}
\label{r}
\langle J_H^0(z)\rangle&=& 8\cos^2 2\theta\:{\rm Im}[\dot{f}(z)]\:I(\beta,m_h,\Gamma),\nonumber\\
I(\beta,m_h,\Gamma)&=&
\int_0^{\infty} du\:\int\frac{d^3 k}{(2\pi)^3}\:\frac{{\rm e}^{-2\Gamma u}}{2\omega_k^2}\frac{\sin^2\omega_k u\:\sin\beta\Gamma}{\left[
{\rm cosh}\beta\omega_k-\cos\beta\Gamma\right]},\nonumber\\
f(z)&=&\left\{2(\lambda_6 c_\theta^2+\lambda_7 s_\theta^2)v_1(z)v_2(z)+2
c_\theta s_\theta\left[\lambda_6 v_1^2(z)+\lambda_7 v_2^2(z)\right]\right\}.
\end{eqnarray}
In the expression above we have only   included  the dominant contribution coming from the diagrams where only nearly Higgs massless modes at the critical temperature propagate; 
$\omega_k^2=k^2+m_h^2$, where $m_h$ denote the mass of the degrees of freedom propagating in the loop whose decay rate in the plasma is $\Gamma$; $\beta=T^{-1}$. It is easy to show that the contribution proportional to $\lambda_5$ is suppressed because $|\lambda_5|\ll |\lambda_{6,7}|$  and because heavy states must propagate in the relative one-loop diagrams. The integral over the `` time '' variable $u$ makes evident the causality
inherent to the nonequilibrium approach: only the processes taking places at times smaller than $t_z$ may give rise to a nonvanishing current. 
 
Performing the integrals in $I(\beta,m_h,\Gamma)$ and taking the limit $\Gamma\ll T$, we obtain the final expression
\begin{equation}
\label{re}
\langle J_H^0(z)\rangle\simeq  \frac{1}{\pi}\cos^2 2\theta\:{\rm Im}[\dot{f}(z)]\frac{T}{\sqrt{ m_h^2+\frac{\Gamma^2}{4}}}.
\end{equation}
Eqs. (\ref{r}) and (\ref{re})  warrant
 some comments. First, we notice that the
momentum integration is
infrared dominated: quasiparticles with long wavelengths and
momentum  perpendicular to the wall give a large contribution to
$\gamma_H(z)$ and a classical approximation is not adequate to
describe the quantum interference nature of $CP$-violation.
Secondly, the final result has the same  infrared singularity responsible for the failure of the perturbative expansion \cite{dj} for the values of the parameters
such that $m_h=0$ (here the singular behaviour is smeared out by the presence of the additional `` mass '' term $\sim \Gamma/2$). However, the theory remain perturbative as far as $\alpha_h\equiv (g^2/2\pi)(T^2/m^2_h)\lsim 1$, which gives $m_h\gsim 0.3\: T$. 
An estimate
of the Higgs damping rate in the SM was obtained in Ref.
\cite{EEV} in the low momentum limit  and can be used here only to
give
very crude estimate of the Higgs coherent time, $\tau\sim
(10-10^{2})/T$. With such values, our derivative expansion is 
barely justified since the wall thickness can certainly span the same range
$(10-10^2)/T$. We will discuss the validity of our approximation in the next section.

The next step amounts to solving the set of coupled differential equations describing the effects of diffusion, particle number changing reactions and CP-violating source terms. We suppose that,  among the supersymmetric particles,  charginos, neutralinos and the right-handed stops as well as the Higgs degrees of freedom,  are in equilibrium in the thermal bath. Under this hypothesis,  strong sphalerons do not drive the asymmetry to zero \cite{giudice}.

Closely following the approach taken in Ref. \cite{nelson,noi}  we can estimate the final baryon asymmetry generated by the Higgs sector to be
\begin{equation}
\left(\frac{n_B}{s}\right)_H=-g(k_i)\frac{{\cal A}\overline{D}\Gamma_{{\rm ws}}}
{v_w^2 s},
\end{equation}
where $s=2\pi^2 g_{*s}T^3/45$ is the entropy density ($g_{*s}$
being
the effective number of relativistic degrees of freedom); 
$g(k_i)$
is a numerical coefficient depending upon the light degrees of
freedom present in the thermal bath, $\overline{D}$ is the effective diffusion constant, $\Gamma_{ws}=6\kappa \alpha_W^4 T$ is the weak sphaleron rate ($\kappa\simeq 1$) \cite{amb}\footnote{The correct value of $\kappa$ is at present the subject of debate, see, for instance, Ref. \cite{ar}.} and
\begin{equation}
\label{higgs2}
{\cal A}= {\cal B}_{+}\left(1-\frac{\lambda_-}{\lambda_+}\right)=
{\cal B}_{-}\left(\frac{\lambda_+}{\lambda_-}-1\right)=
\frac{1}{\overline{D} \; \lambda_{+}} \int_0^{\infty} du\;
\widetilde \gamma_H(u)
e^{-\lambda_+ u},
\end{equation}
where 
\begin{equation}
\lambda_{\pm} = \frac{ v_w \pm
\sqrt{v_w^2 + 4 \widetilde{\Gamma}
\overline{D}}}{2 \overline{D}}.
\end{equation}
The quantity $\widetilde{\Gamma}$ is the effective decay constant and  $\widetilde \gamma_H({\bf z})= v_w \partial_{{\bf z}}
J_H^0({\bf z}) f(k_i)$ is now  defined in the bubble wall frame, 
$f(k_i)$ being   a coefficient depending on the number of
degrees of freedom present in
the thermal bath and related to the definition of the
effective source~\cite{nelson}. Since for relatively low  values of  the pseudoscalar mass $m_A$ ($m_A\gsim m_Z$) the variation of the ratio of the vacuum expectation values of the Higgs fields along the bubble wall is  small \cite{early3,iiro}, we can take $\theta\sim\beta$. For typical values $\tan\beta\sim 2$, $\Gamma\simeq 5\times 10^{-2} T$, $m_h\simeq 0.4\: T$ and $\sqrt{v_1^2(T)+v_2^2(T)}/T\simeq 1.2$, we find 
\begin{equation}
\left(\frac{n_B}{s}\right)_H\simeq -6\times 10^{-11}\sin\phi_\mu,
\end{equation} 
where the dependence upon $v_w$ is very weak.

The Higg scalar sector may account for the baryon asymmetry for large values of the phase $\phi_\mu$. Such large values may be tolerated in view of the experimental limits on the neutron electric dipole moment of the neutron if the squarks of the first and second generation have masses of the order of a few TeV.

{\bf 3.}~~One should not claim victory too soon, though. Let us look back at the approximations we have made and discuss to which extent  our estimate might be altered by performing a more refined analysis:

{\it a)} Our result is rigously valid only for $L_w\Gamma\gsim 1$. Given our poor knowledge of both parameters, this condition does not seem unreasonable. Moreover, for $L_w\Gamma\ll 1$ one may expect that the dependence of the final result upon $\Gamma$ disappears \cite{nelson} so that $I(\beta,m_h,\Gamma)$ will saturate at the value $\Gamma\sim L_w^{-1}$. Our results seem to confirm this expectation.

{\it b)} We have seen that quasiparticles with long wavelengths  give the  largest contribution to the source. This means that the classical approximation is not adequate to
describe the quantum interference nature of $CP$-violation and a quantum approach must be adopted to compute the CP-violating source. This was done in the present paper. However, in the second stage of the computation, we have made use of classical Boltzmann equations. For low momentum particles, 
the validity of the classical Boltzmann equation  starts to break down. It is indisputable that the ultimate answer can
be provided only by a complete nonequilibrium quantum field theory
approach. Kinetic theory and classical Boltzmann equations have been used to
describe the dynamics of particles treated as classical with a defined
position, energy and momentum.  This requires that, in
particular, the mean free path must be large compared to the Compton
wavelength of the underlying particle in order for the classical picture
to be valid, which is not guaranteed for
particles with a small momentum perpendicular to the
wall. Distribution functions obeying the quantum Boltzmann equations are the  only  correct
functions to describe particles in an interacting, many-particle
environment. 

Solving the quantum Boltzmann equations represents an 
Herculean task. However, recent investigations have revealed that the quantum Wigner distributions posses strong memory effects and that their relaxation time is  typically longer than the one obtained in the classical limit \cite{henning}. This slowdown of the relaxation processes  may keep the system out of equilibrium for longer times and therefore enhance the final baryon asymmetry. 

{\it c)}  When computing the Higgs source, we have been assuming  the loopwise expansion being reliable, in other words we have computed only the lowest order loop diagrams. However, there is an infinite class of loop diagrams with more interaction vertices, the so-called `` ladder '' diagrams \cite{jeon}, which may contribute at the leading order even though
full thermal corrections have been used in the evaluation of the lowest level diagrams.  This is because there are additional infrared divergencies at finite
temperature when the ladder-diagrams contain  nearly  massless Higgs modes and the infrared cut-off is provided by  the mass $m_h$. The physical reason for this large correction has to be identified with the fact that the lowest loop diagram takes into account only the effects of particles which have undergo only a few collisions in the plasma, after which particles are not fully thermalized yet.  Summing up all the leading contributions is a nontrivial problem (in fact, it has been recently claim that the ladder graphs may be canceled by other types of terms in the loop expansion \cite{ladder}). A reasonable estimate  of the relative size of of the contribution of the ladder-diagrams with respect to the lowest  diagram computed in this paper is given by (loop suppression factor)$\times (g^2 T^2/m_h^2)\ln(T/m_h)$. Numerically this contribution may be sizeable and might change the final estimate of the baryon asymmetry by some factor ${\cal O}(1)$.

\vskip 1cm
\underline{Acknowledgements}:

It is a plasure to thank my collaborators  M. Carena, M. Quiros,  I. Vilja and C.E.M. Wagner for useful discussions. 

\def\NPB#1#2#3{Nucl. Phys. {\bf B#1}, #3 (19#2)}
\def\PLB#1#2#3{Phys. Lett. {\bf B#1}, #3 (19#2) }
\def\PLBold#1#2#3{Phys. Lett. {\bf#1B} (19#2) #3}
\def\PRD#1#2#3{Phys. Rev. {\bf D#1}, #3 (19#2) }
\def\PRL#1#2#3{Phys. Rev. Lett. {\bf#1} (19#2) #3}
\def\PRT#1#2#3{Phys. Rep. {\bf#1} (19#2) #3}
\def\ARAA#1#2#3{Ann. Rev. Astron. Astrophys. {\bf#1} (19#2) #3}
\def\ARNP#1#2#3{Ann. Rev. Nucl. Part. Sci. {\bf#1} (19#2) #3}
\def\MPL#1#2#3{Mod. Phys. Lett. {\bf #1} (19#2) #3}
\def\ZPC#1#2#3{Zeit. f\"ur Physik {\bf C#1} (19#2) #3}
\def\APJ#1#2#3{Ap. J. {\bf #1} (19#2) #3}
\def\AP#1#2#3{{Ann. Phys. } {\bf #1} (19#2) #3}
\def\RMP#1#2#3{{Rev. Mod. Phys. } {\bf #1} (19#2) #3}
\def\CMP#1#2#3{{Comm. Math. Phys. } {\bf #1} (19#2) #3}

\end{document}